\documentclass[12pt]{article}

\usepackage{amsmath}
\usepackage{graphicx,psfrag,epsf}
\usepackage{times}
\usepackage{bm}
\usepackage{natbib}

\usepackage[plain,noend]{algorithm2e}

\newcommand{\blind}{1}

\usepackage{bbm}
\usepackage{amssymb}

\newtheorem{example}{Example}
\newtheorem{theorem}{Theorem}

\def\cov{\hbox{cov}}
\def\log{\hbox{log}}

\def\bq{\begin{equation}}
\def\eq{\end{equation}}

\def\trans{^{ \mathrm{\scriptscriptstyle T} }}

\begin{document}

\def\spacingset#1{\renewcommand{\baselinestretch}%
	{#1}\small\normalsize} \spacingset{1}


\if1\blind
{
	\title{\bf Hierarchical low rank approximation of likelihoods for large spatial
		datasets}
	\author{Huang Huang and Ying Sun \hspace{.2cm}\\
		CEMSE Division, \\King Abdullah University of Science and Technology, 
		\\Thuwal, 23955, Saudi Arabia.\\
		}
	\date{Jan. 19, 2016}
	\maketitle
} \fi

\if0\blind
{
	\bigskip
	\bigskip
	\bigskip
	\begin{center}
		{\LARGE\bf Hierarchical low rank approximation of likelihoods for large spatial
			datasets}
	\end{center}
	\medskip
} \fi

\bigskip
\begin{abstract}
	Datasets in the fields of climate and environment are often very large and irregularly spaced. To model such datasets, the widely used Gaussian process models in spatial statistics face tremendous challenges due to the prohibitive computational burden. Various approximation methods have been introduced to reduce the computational
	cost. However, most of them rely on unrealistic assumptions for the
	underlying process and retaining statistical efficiency remains an
	issue. We develop a new approximation scheme for maximum likelihood estimation.
	We show how the composite likelihood method can be
	adapted to provide different types of hierarchical low rank
	approximations that are both computationally and statistically efficient. The improvement of the proposed method is explored theoretically;
	the performance is investigated by numerical and simulation studies; and the practicality is illustrated through applying our methods to 2 million measurements of soil moisture in the area of the Mississippi River basin, which facilitates a better understanding of the climate variability.
\end{abstract}

\noindent%
\vfill

\newpage
\spacingset{1.45} 


\section{Introduction}\label{sec:intro}
Soil moisture is a key factor in climate systems, which has a significant impact on hydrological processes, runoff generations and drought developments. To understand its spatial variability and predict values at unsampled locations, Gaussian process models are widely used \citep{Stein:1999}, where likelihood based methods are appropriate for model fitting. However, it generally requires $O(n^3)$ computations and $O(n^2)$ memory for $n$ irregularly spaced locations \citep{Sun:Stein:2014}. Similar to other climate variables, many satellite-based or numerical model generated soil moisture datasets have nearly a global coverage with high spatial resolutions, so that the exact computation of Gaussian likelihood becomes prohibitive. There are various existing methods, many of which were discussed by \citet{Sun:Li:Genton:2012}.  For example, 
covariance tapering \citep{Furrer:Genton:Nychka:2006,Kaufman:Schervish:Nychka:2008,Sang:Huang:2012} assumes a compactly supported covariance function, which leads to a sparse covariance matrix; low rank models, including 
space-time Kalman filtering \citep{Wikle:Cressie:1999},
low rank splines \citep{Lin:Wahba:Xiang:Gao:Klein:Klein:2000},
moving averages \citep{Hoef:Cressie:Barry:2004},
predictive processes \citep{Banerjee:Gelfand:Finley:Sang:2008} and 
fixed rank kriging \citep{Cressie:Johannesson:2008}, make use of a latent process with a lower dimension where the resulting covariance matrix has a low rank representation; and Markov random field models \citep{Cressie:1993,Rue:Tjelmeland:2002,Rue:Held:2005,Lindgren:Rue:Lindstrom:2011} exploit fast-approximated conditional distributions assuming conditional independence with the precision matrix being sparse. These methods use models that may allow exact computations to reduce computations and/or storage, and each has its strength and weakness. For instance, \citet{Stein:2013} studied the properties of the covariance tapers and showed that covariance tapering sometimes performs even worse than assuming independent blocks in the covariance; \citet{Stein:2014} discussed the limitations on the low rank approximations; and Markov models  depend on the observation locations, and realignment to a much finer grid with missing values is required for irregular locations \citep{Sun:Stein:2014}. Recently developed methods include the
nearest-neighbor Gaussian process model \citep{Datta:Banerjee:Finley:Gelfand:2015}, which 
is used as a sparsity-inducing prior within a Bayesian hierarchical modeling framework,
the multiresolution Gaussian process model \citep{Nychka:Bandyopadhyay:Hammerling:Lindgren:Sain:2015}, which constructs basis functions using compactly supported correlation function on different level of grids,
equivalent kriging \citep{Kleiber:Nychka:2015}, which uses an equivalent kernel to approximate the kriging weight function when a nontrivial nugget exists, and multi-level restricted Gaussian maximum likelihood method \citep{Candas:Genton:Yokota}, for estimating the covariance function parameters using contrasts.

An alternative way to reduce computations is via likelihood and score equation approximations. \citet{Vecchia:1988} first proposed to approximate the likelihood using the composite likelihood method, where the conditional densities were calculated by choosing only a subset of the complete conditioning set. \citet{Stein:Chi:Wetly:2004} adapted this method for restricted maximum likelihoods approximation. Instead of approximating the likelihood itself, \citet{Sun:Stein:2014} proposed new unbiased estimating equations for score equation approximation, where the sparse precision matrix approximation is constructed by a similar method. In these approximation methods, the exact likelihood and the score equations can be obtained by using the complete conditioning set to calculate each conditional density. It was shown that the approximation quality or the statistical efficiency depends on the selected size of the subset. It is common that the subset is still inadequate by considering the largest possible number of nearest neighbors, which motivates this work.

In this paper, we propose a generalized hierarchical low rank method for likelihood approximation. The proposed method utilizes low rank approximations hierarchically, which does not lead to a low rank covariance matrix approximation. Therefore, it is different from the predictive process method \citep{Banerjee:Gelfand:Finley:Sang:2008}, where the covariance matrix is approximated by a low rank representation. Furthermore, the proposed method contains the independent blocks \citep{Stein:2013} and nearest neighbors \citep{Sun:Stein:2014} approaches as special cases. The improvement of the proposed method is explored theoretically and the performance is investigated by numerical and simulation studies. We show that the hierarchical low rank approximation significantly improves the statistical efficiency of the most commonly used methods while retaining the computational efficiency, especially when the size of conditional subsets is restricted by the computational capacity, which is always the case for real datasets. For illustrations, our method is applied to a large real-world spatial dataset of soil moisture in the Mississippi River basin, U.S.A., to facilitate a better understanding of the hydrological process and climate variability. Our method is able to fit a Gaussian process model to 2 million measurements with fast computations, making it practical and attractive for very large datasets.

\section{Methodology}\label{sec:meth}

\subsection{Approximating likelihoods}\label{subsec:likeli}

Let $\{z(s):s\in
D\subset\mathbb{R}^d\}$ be a stationary isotropic Gaussian Process in a domain $D$
in the $d$-dimensional Euclidean space, and typically $d=2$. 
We assume the mean of the process is zero for simplicity and the covariance function has a parametric form $C(h;\theta)=\cov\{z(s),z(s')\}$, where $h=\|s-s'\|$ and $\theta$ is the parameter vector of length $p$. 
Suppose that data are observed at $n$ irregularly spaced locations $s_1,\ldots,s_n$, then,
\[
Z=(z_1,\ldots,z_n)\trans\sim N(0,\Sigma(\theta)),
\]
where $z_i=z(s_i), i=1,\ldots,n$, and $\Sigma(\theta)$ is the variance-covariance matrix with the $(i,j)^\textrm{th}$ element $C(\|s_i-s_j\|;\theta).$ For simplicity, $\theta$ is omitted in notations hereinafter unless clarification is needed.

The maximum likelihood estimate can be obtained by maximizing the log-likelihood, 
\[
\ell(\theta\mid Z)=\log\{f(Z\mid\theta)\}=-\frac{1}{2}\log(|\Sigma|)
-\frac{1}{2}Z\trans\Sigma^{-1}Z-\frac{n}{2}\log(2\pi),
\]
where $f$ is the multivariate normal density. In practice, if the mean of $Z$ is a vector that depends linearly on unknown parameters, the restricted maximum likelihood estimate should be employed \citep{Stein:Chi:Wetly:2004}.

When computations become prohibitive, one way to approximate the likelihood is through log-conditional densities,
\[
\ell(\theta\mid Z)=\log\{f(z_1\mid\theta)\}+\sum_{j=1}^{n-1}\log\{
f(z_{j+1}\mid Z_j,\theta)\},
\]
where $Z_{j}=(z_1,\ldots,z_{j})\trans,$ for $1\leq j\leq n-1,$ indicating all the ``past" observations of $z_{j+1}$. 
Since,
\[
\cov\Biggl(
\begin{array}{c}
Z_{j} \\
z_{j+1} \\
\end{array}
\Biggr)=\Biggl(
\begin{array}{cc}
\Sigma_{jj} & \sigma_{j} \\
\sigma_{j}\trans & \sigma_{j+1,j+1} \\
\end{array}
\Biggr),
\]
it is easy to show that for $j=1,\ldots,n-1$,
\bq\label{eq:cond}
\log \{f(z_{j+1}|Z_{j})\}=-\frac{1}{2}\Biggl\{\frac{\bigl(z_{j+1}-\sigma\trans_{j}\Sigma_{jj}^{-1}Z_{j}\bigr)^2}{\sigma_{j+1,j+1}-\sigma\trans_{j}\Sigma_{jj}^{-1}\sigma_{j}}
+\log\bigl(\sigma_{j+1,j+1}-\sigma\trans_{j}\Sigma_{jj}^{-1}\sigma_{j}\bigr)
+\log(2\pi)\Biggr\},
\eq
which is the log-density of $w_j=b_j\trans Z,$ where $b_j=(-\sigma_{j}\trans\Sigma_{jj}^{-1},1,0,\ldots,0)\trans.$
It can be shown that $w_j$s are independent and $w_j\sim N(0,v_j),$ where $v_j=b_j\trans\Sigma b_j$ \citep{Stein:Chi:Wetly:2004}. \citet{Sun:Stein:2014} further showed that the precision matrix is $\Sigma^{-1}=\sum_{j=0}^{n-1}b_jb_j\trans/v_j$, where $b_0=(1,0,\ldots,0)\trans$ and $v_0=b_0\trans\Sigma b_0$.

More generally, $z_{j+1}$ can be defined as a vector which is usually more computationally efficient, and the corresponding $b_j=(-\sigma_{j}\trans\Sigma_{jj}^{-1},I,0,\ldots,0)\trans,$ where $I$ is an identity matrix of size $j$. 

However, for a large $j$, it is computationally expensive to evaluate $\Sigma_{jj}^{-1}\sigma_{j}.$ \citet{Vecchia:1988} proposed approximating each conditional density by only conditioning on a subset $z_{j+1}$ consisting of $r\ll j$ nearest neighbors. The same approach is used by \citet{Stein:Chi:Wetly:2004}
for approximating the restricted maximum likelihood estimate. \citet{Sun:Stein:2014} also used the subset of nearest neighbors to approximate the precision matrix for score equation approximation.

In this paper, we propose a generalized framework that allows to approximate these conditional densities hierarchically using a low rank representation. Although we implement our algorithm for application in \S\ref{sec:app} with $z_{j+1}$ being a vector, we present and illustrate our methodology assuming $z_{j+1}$ is scalar for simplicity.

\subsection{Hierarchical low rank representation}\label{subsec:approx}
Motivated by the nearest neighbors method, where only $r\ll j$ nearest neighbors are selected to approximate $\Sigma^{-1}_{jj}\sigma_{j}$ for a large $j$ in equation~\eqref{eq:cond}, we propose a general approximation framework for $j>r$ using a low rank representation.

Denote $\Sigma^{-1}_{jj}\sigma_{j}$ by $x_j,$ or $\Sigma_{jj}x_j=\sigma_j$.
We propose to approximate $x_j$ by a low rank representation $\hat x_j=A_{j,r}\tilde x_j,$ where $\tilde x_j$ is a vector of length $r$ and $A_{j,r}$ is a $j\times r$ matrix. Then, instead of solving $\Sigma_jx_j=\sigma_j$, we minimize the norm $\|\Sigma_{jj}A_{j,r}\tilde x_j-\sigma_j\|_{\Sigma{_{jj}^{-1}}}=(\Sigma_{jj}A_{j,r}\tilde x_j-\sigma_j)\trans\Sigma_{jj}^{-1}(\Sigma_{jj}A_{j,r}\tilde x_j-\sigma_j)$ or equivalently solve $A_{j,r}\trans\Sigma_{jj}A_{j,r}\tilde x_j=A_{j,r}\trans\sigma_j.$ Therefore, $x_j$ is approximated by,
\bq\label{eq:approx}
\hat x_j=A_{j,r}\tilde x_j=A_{j,r}(A_{j,r}\trans \Sigma_{jj}A_{j,r})^{-1}A_{j,r}\trans\sigma_{j},
\eq
which only involves a linear solve of dimension $r$. 
In this framework, we approximate $x_j$ for each $j>r$ hierarchically by a low rank representation,
which includes many commonly used strategies as special cases with different choices of $A_{j,r}$. The following are some examples:

\begin{example}
	Independent blocks method (IND). In this
	method, no correlation between ``past" points and the ``current" point is
	considered. Namely, $A_{j,r}$ is a $0$ matrix; however, $z_{j+1}$ is a vector of length $r$ here for fair comparison to other methods in terms of computation.
\end{example}

\begin{example}
	Nearest neighbors method (NN). Choose $r$ nearest neighbors of $z_{j+1}$ from $Z_j$. The corresponding $A_{j,r}$ is of $j\times r$ dimensions, where each column consists of only one element $1$ at the $k$-th row if $z_k$ is selected from $Z_j$ and zero otherwise.
\end{example}

\begin{example}
	Nearest neighboring sets method (SUM). Choose $r$ nearest neighboring sets of $z_{j+1}$, where each set contains $m>1$ neighbors and a total of $mr\ll j$ neighbors are selected from $Z_j$. The matrix $A_{j,r}$ is specified as a $j\times r$ matrix with each column having $m$ elements of $1$, indicating the sum of the $m$ selected neighbors are considered. In this way, more neighbors are included while the computational cost remains the same.
\end{example}

\begin{example}
	Nearest neighbors and nearest neighboring sets method (NNSUM). Combine Examples~2 and 3, where $r_1$ columns of $A_{j,r}$ are constructed as in Example~2, and $r-r_1$ are built as in Example~3. In this way, we use the exact information from the $r_1$ nearest neighbors and consider $r-r_1$ nearest neighboring sets with a total number of $r_1+m(r-r_1)$ selected nearest neighbors.
\end{example}

\subsection{Hierarchical low rank approximation method}\label{subsec:HLR}
In this section, we propose a generalized hierarchical low rank approximation method (HLR). In equation~\eqref{eq:approx}, the matrix $A_{j,r}$ is a $0$-$1$ matrix. The $r\times r$ matrix $A_{j,r}\trans\Sigma_{jj}A_{j,r}$ only extracts the corresponding rows or columns of $\Sigma_{jj}$. Now suppose we select $mr$ nearest neighbors of $z_{j+1}$, and the corresponding $A_{j,mr}$ is of size $j\times mr.$ To retain the same computational costs associated with rank $r$, we propose the following approximation,
\bq
\label{eq:nugget}
A_{j,mr}\trans\Sigma_{jj}A_{j,mr}\approx P_jL_{j}P_j\trans+\epsilon_j^2I_{mr},
\eq
where $L_{j}$ is a positive definite matrix of dimension $r\times r,$ $P_j$ is a 
$mr\times r$ matrix consisting of $r$ basis functions, $I_{mr}$ is the identity matrix of size $mr$, and $\epsilon_j^2$ is the
nugget. 
By the Sherman--Morrison--Woodbury formula,
\bq\label{eq:HLR}
(P_jL_{j}P_j\trans+\epsilon_j^2I_{mr})^{-1}= \epsilon_j^{-2}I_{mr}-\epsilon_j^{-4}P_j(L_{j}^{-1}+\epsilon_j^{-2}P_j\trans P_j)^{-1}P_j\trans,
\eq
then $(A_{j,mr}\trans\Sigma_{jj}A_{j,mr})^{-1}$ in equation \eqref{eq:approx} can be approximated by inverting only an $r\times r$ matrix $L_{j}$. This approach shares the same spirits with the predictive process \citep{Banerjee:Gelfand:Finley:Sang:2008} and fixed rank kriging \citep{Cressie:Johannesson:2008}.
However both methods approximate the covariance function by a low rank representation while the low rank approximation is done for each $j>r$ hierarchy in our method, and the resulting approximated covariance is no longer low rank. The detailed choice for $P_j$ were discussed by \citet{Cressie:Johannesson:2008}. In this paper, we use the eigenfunctions.

\begin{figure}[b!]
	\centering
	\includegraphics[width=0.8\textwidth]{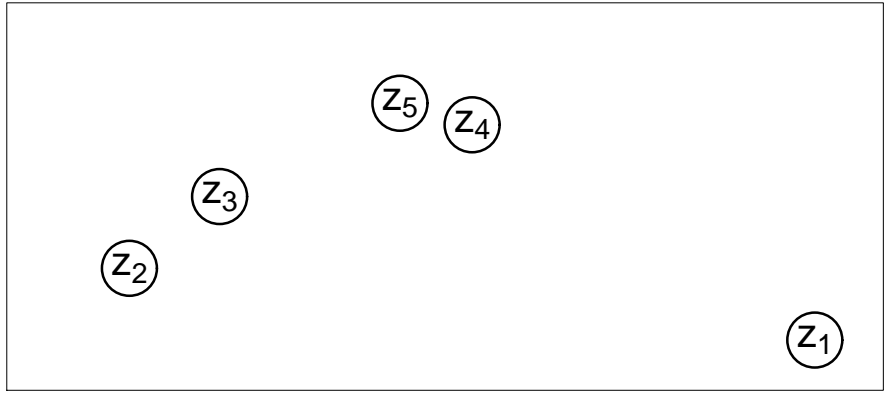}
	\caption{\label{fig:method}A random field where $n=5$ locations have observations}.
\end{figure}

To help comprehend, Fig.~\ref{fig:method} illustrates the methods described in \S\ref{subsec:approx} and \ref{subsec:HLR} for $n=5$ observations $Z=(z_1,\ldots,z_5)\trans$. Let $r=2$, then IND considers $3$ independent blocks, and $f(Z)$ is approximated by 
$f(z_5)f(z_4,z_3)f(z_2,z_1)$. For the other four methods, the conditional density is required to calculate in each hierarchy. For instance, in hierarchy $j=5$, NN approximates the conditional density $f(z_5\mid z_4,\ldots,z_1)$ by $f(z_5\mid z_4,z_3)$; SUM by $f(z_5\mid z_4+z_3,z_2+z_1)$; NNSUM by $f(z_5\mid z_4,z_3+z_2)$; and HLR by
$f(z_5\mid a_{14}z_4+a_{13}z_3+a_{12}z_2+a_{11}z_1,
a_{24}z_4+a_{23}z_3+a_{22}z_2+a_{21}z_1)$, where $a_{ij}$s are determined by the low rank approximation.

\subsection{Assessing model quality}\label{subsec:quality}
There are various ways to measure the performance of approximation methods, including the Kullback--Leibler divergence, the Godambe information matrix, and the Frobenius norm. 

The Kullback--Leibler divergence computes the divergence of the approximated from the exact distributions. 
For the zero-mean Gaussian process, the Kullback--Leibler divergence has the closed form,
\[
D_\textsc{K-L}(N_e\|N_a)=\frac{1}{2}\left\{\hbox{tr}(\Sigma_a^{-1}\Sigma_e)+\log(|\Sigma_a|)
-\log(|\Sigma_e|)-n\right\},
\]
where $N_e$ and $N_a$ stand for the exact and the approximated distributions, respectively, $\Sigma_e$ and $\Sigma_a$ are the corresponding covariance matrices, and $n$ is the dimension of the distribution.

The Godambe information matrix gives the asymptotic variances and covariances for the estimated parameters in the Gaussian process, as used by \citet{Kaufman:Schervish:Nychka:2008} and \citet{Sun:Stein:2014}.
The Frobenius norm is another way to think about this problem. However, it is a matrix norm and does not penalize the positive definiteness of a covariance matrix~\citep{Stein:2014}.

For our numerical and simulation studies in \S3, we choose the Kullback--Leibler divergence and the Godambe Information matrix to assess the quality of the approximation. Because the results in terms of showing the different performances are similar, we only present the results of Kullback--Leibler divergence. It will be shown numerically that the Kullback--Leibler divergence of the hierarchical low rank approximation method is always the smallest. 
This is due to the fact that for each $j>r$, the hierarchical low rank approximation method provides a better approximation in equation~\eqref{eq:approx} by including more neighbors than the nearest neighbors method. Let $V^\textsc{N}_{jj}$ be the $r\times r$ matrix defined by $A_{j,r}\trans \Sigma_{jj}A_{j,r}$ in equation~\eqref{eq:approx} using the nearest neighbors method and let $V^\textsc{H}_{jj}=P_jL_{j}P_j\trans+\epsilon_j^2I_{mr}$ be the $mr\times mr$ matrix for approximating $A_{j,mr}\trans \Sigma_{jj}A_{j,mr}$ in equation~\eqref{eq:nugget} by the hierarchical low rank approximation method, where $P_j$ consists of eigenfunctions. The following theorem shows the result that the approximation to $\Sigma_{jj}$ induced by $V^\textsc{H}_{jj}$ is better than that induced by $V^\textsc{N}_{jj}$ in terms of the Frobenius norm.

\begin{theorem}
	\label{theorem}
	 Let $\lambda_1\geq\lambda_2\geq\cdots\geq\lambda_{mr}>0$ be the eigenvalues of $A_{j,mr}\trans \Sigma_{jj}A_{j,mr}$. If $\epsilon_j^2$ in equation~\eqref{eq:nugget} satisfies 
	 $\epsilon_j^2<(\lambda_r+\lambda_{mr})/2$,
	 we have,
	\[
	\|A_{j,mr}V^\textsc{H}_{jj}A_{j,mr}\trans-\Sigma_{jj}\|_F
	\leq
	\|A_{j,r}V^\textsc{N}_{jj}A_{j,r}\trans-\Sigma_{jj}\|_F,
	\]
	where $\|\cdot\|_F$ means the Frobenius norm.
\end{theorem}
The proof is shown in the Appendix. Similar results hold for the comparison between hierarchical low rank approximation method and the nearest neighboring sets method, or the nearest neighbors and nearest neighboring sets method.

\subsection{Computational complexity and parallelization}\label{subsec:comp}
For our hierarchical low rank approximation method, we need to execute a linear solve of dimension $r$, which requires $O(\min(j,r)^3)$ computation in equation~\eqref{eq:HLR} for each hierarchy $j=1,\ldots,n-1$ assuming that the direct method is employed. Then the total computational cost is $O(r^3n)$ for likelihood approximation per value. When $r\ll n$, the computational cost is much smaller than $O(n^3)$, which is required by the Cholesky decomposition.

In practice, the computation time can be reduced further by choosing $z_{j+1}$ as a vector due to the fact that it leads to a smaller number of hierarchies that need to be evaluated. 
It is also worth noting that our approach can be parallelized easily because the computation of each hierarchy is independent of each other.

\section{Numerical study}\label{sec:num}
\subsection{Design setup}
In the numerical study in this section and the following simulation study in \S\ref{sec:sim}, we focus on irregularly
spaced data with an unstructured covariance matrix \citep{Sun:Stein:2014}. The observations are generated
at the locations $n^{-1/2}(r-0.5 + X_{r\ell},\ell-0.5 +
Y_{r\ell})$ 
for $r,\ell\in\{1,\ldots,n^{1/2}\}$, where $n$ is the number of locations, and $X_{r\ell}$s and
$Y_{r\ell}$s are independent and identically distributed, uniform on $(-0.4, 0.4).$ 
The advantage of this design is that it is irregular, and we can guarantee that no two locations are too close.

Here, 
we study the performances of different approximation methods proposed
in \S\ref{subsec:approx} and \S\ref{subsec:HLR} in different settings.
We consider a zero-mean Gaussian process model with Mat\'ern covariance function possibly with a nugget,
\bq\label{eq:matern}
C(h;\alpha,\beta,\nu,\tau^2)=
\alpha
\{(2\nu)^{1/2}h/\beta\}^\nu
K_\nu\{{(2\nu)^{1/2}}h/\beta\}
/
\{\Gamma(\nu)2^{\nu-1}\}+\tau^2\mathbbm{1}(h=0),
\eq
where $K_\nu(\cdot)$ is the modified Bessel function of the second kind of order $\nu$, $\Gamma(\cdot)$ is the gamma function, $\mathbbm{1}(\cdot)$ is the indicator function, $h\geq 0$ is the distance between two locations, $\alpha>0$ is the sill parameter, $\beta>0$ is the range parameter, $\nu>0$ is the smoothness parameter, and $\tau^2$ is the nugget effect.

For $n$ irregularly spaced locations, the description of the five methods considered is shown in Table~\ref{tab:method}.

\begin{table}[h]
	\centering
	\def~{\hphantom{0}}
	\caption{Description of the five methods used in the numerical study. IND, independent blocks method; NN, nearest neighbors method; SUM, nearest neighboring sets method; NNSUM, nearest neighbors and nearest neighboring sets method; HLR, the hierarchical low rank approximation method.}
	
\begin{tabular}[c]{l|p{.7\textwidth}}
	\hline
	Method&Description\\\hline
	IND& Divide the locations into $\lceil n/r\rceil$ blocks, each of which contains $r$ points. $\lceil n/r \rceil$ means the largest integer that is no larger than $n/r$.
	\\
	NN& A number of $r$ nearest neighbors are selected to construct $A_{j,r}.$
	\\
	SUM& A number of $r$ nearest neighboring sets are selected and each set has 2 locations. Then a total number of $2r$ nearest neighbors are used to construct $A_{j,r}$.
	\\
	NNSUM& A number of $\lceil
	r/2\rceil$ nearest neighbors are first selected, then the following $2(r-\lceil
	r/2\rceil)$ nearest neighbors are divided into $r-\lceil
	r/2\rceil$ sets of size $2$.
	\\
	HLR& A number of $2r$ nearest neighbors are considered, where $L_{j}$ is a $r\times r$ diagonal matrix with elements corresponding to the $r$ leading eigenvalues. $P$ consists of the $r$ corresponding eigenvectors.
	\\
	\hline
\end{tabular}
\label{tab:method}
\end{table}

In \S\ref{subsec:depend}--\ref{subsec:noise}, we present the Kullback--Leibler divergence calculated from different settings for the five methods with $\alpha$ fixed at $1$ and $n=900$. In \S\ref{subsec:further}, we discuss the effect of sample size $n$ and the rank $r$.

\subsection{Dependence level}\label{subsec:depend}
In the Mat\'ern model in equation~\eqref{eq:matern}, the range parameter $\beta$ controls the dependence of the process. In this section, we consider different $\beta$.
Given $\nu=0.5$, which corresponds to an exponential covariance function and $\tau^2=0.15$, the first row of Fig.~\ref{fig:numericalStudy} shows the Kullback--Leibler divergence for $\beta=0.1$, which means a weaker dependence, and $\beta=0.5$, which indicates a stronger dependence, as the rank $r$ increases from $2$ to $8$. We can see that the HLR approximation is always the best with the smallest Kullback--Leibler divergence, and SUM and NNSUM win against NN only when $r=2$ for $\beta=0.1$, while for $\beta=0.5$, the improvement of SUM and NNSUM exists up to $r=6$.
It implies that when a strong correlation is present, a small number of nearest neighbors is not adequate to
provide a good approximation of the conditional density. 
It is also worth noting that the range of $r/n$ in this study is from $0.22\%$ to $0.89\%$. For very large $n$, and $r\ll n$, the improvement from HLR, SUM or NNSUM approaches can be substantial.

\subsection{Smoothness level}
\begin{figure}[p!]
	\centering
	\includegraphics[width=0.8\textwidth]{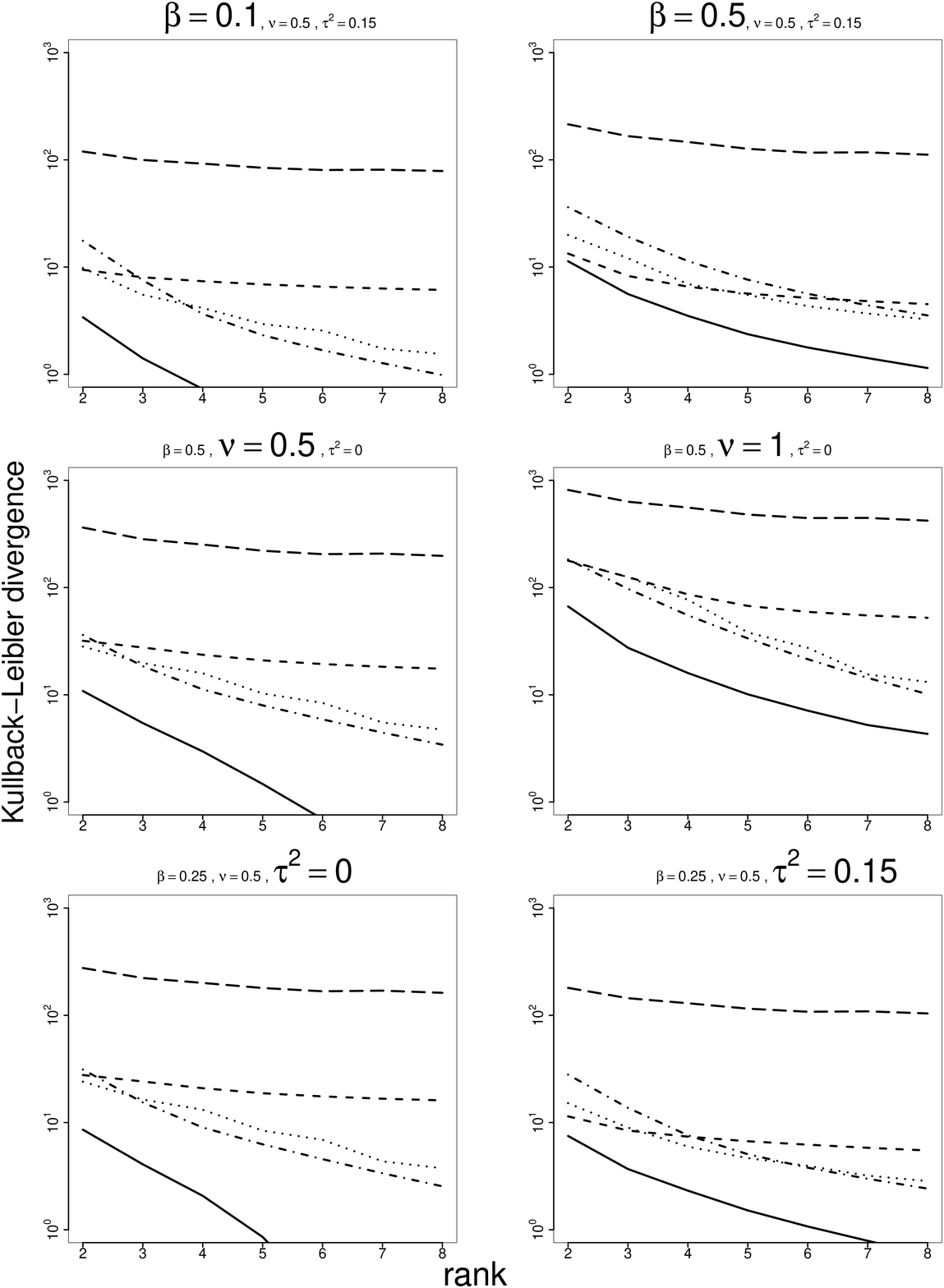}
	\vskip -0.7cm
	\caption{\label{fig:numericalStudy}Six panels showing the Kullback--Leibler divergence against rank with 900 locations in IND (long-dash, $---$), NN (dot-dash, $\cdot$ - $\cdot$), SUM (dashes, - - -), NNSUM (dots, $\cdot\cdot\cdot$), and HLR (solid, ---) methods. The corresponding parameters are indicated in the titles.}
\end{figure}

In the Mat\'ern covariance function, a larger $\nu$ indicates a smoother process.
In this section, we fix $\beta=0.5$ and $\tau^2=0.$ We consider two smoothness levels with $\nu=0.5$ and $\nu=1,$ which correspond to the exponential and Whittle covariance functions, respectively.
The second row of Fig.~\ref{fig:numericalStudy} shows the Kullback--Leibler divergence. 
Similarly, the HLR approach outperforms the other methods. For the rougher process, when $\nu=0.5$, SUM and NNSUM are slightly better than NN at $r=2.$ When $\nu$ increases to $1$, the improvement almost disappears and all the methods need a large $r$ to achieve similar performances as $\nu=0.5.$

\subsection{Noise level}\label{subsec:noise}
The nugget effect can be viewed as measurement errors or the micro-structure in the underlying process. In this section, we consider different $\tau^2$. Given $\beta=0.25$ and $\nu=0.5$, the last row of Fig.~\ref{fig:numericalStudy} shows the Kullback--Leibler divergence for $\tau^2=0$ and $\tau^2=0.15.$ In both cases, the HLR approach still provides the best approximation, although for large $\tau^2,$ a larger $r$ is needed. If the rank $r$ is limited to a small number, we can see that SUM or NNSUM can improve NN when the process is noisy or with a larger $\tau^2.$

\subsection{Sample size and rank}\label{subsec:further}
\begin{figure}[b!]
	\centering
	\includegraphics[width=0.8\textwidth]{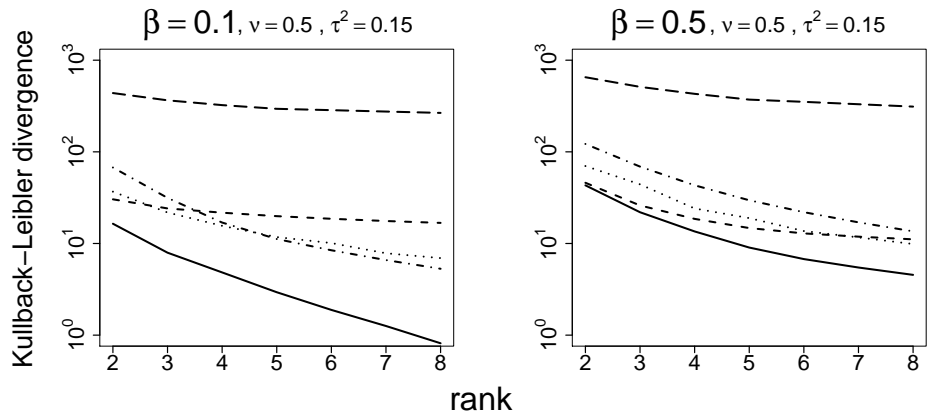}
	\caption{\label{fig:range50}Two panels showing the Kullback--Leibler divergence against rank with 2500 locations in IND (long-dash, $---$), NN (dot-dash, $\cdot$ - $\cdot$), SUM (dashes, - - -), NNSUM (dots, $\cdot\cdot\cdot$), and HLR (solid, ---) methods. The corresponding parameters are indicated in the titles.}
\end{figure}

In this section, we explore the effect of sample size given the rank $r$ or the ratio of $r/n$. Fig.~\ref{fig:range50} shows the results for a similar design as in the first row of Fig.~\ref{fig:numericalStudy} but with $n=2500$.
Comparing Fig.~\ref{fig:range50} to the first row of Fig.~\ref{fig:numericalStudy}, we can see that for a given process, a larger number of locations does require larger ranks to achieve a similar approximation quality. When $r$ is fixed, NN is often not adequate, especially for large $n$, and SUM, NNSUM, and HLR can improve the approximation by including more neighbors.

Although it is not realistic for a large dataset, we also investigate a situation where NN is adequate to provide a good approximation at rank $r$, and then compare the Kullback--Leibler divergence for NN at $r+1$ and NNSUM with the same first $r$ nearest neighbors and one additional set containing the next $2$ nearest neighbors. We find that for $\alpha=1, \beta=0.5, \nu=0.5, \tau^2=0$ and $n=900$, NN with rank $r+1=51$ gives a Kullback--Leibler divergence as $9.4\times10^{-2}$ and NNSUM reduces Kullback--Leibler divergence by $1\%.$

\section{Simulation study}\label{sec:sim}
\begin{figure}[h!]
	\centering
	\includegraphics[width=0.8\textwidth]{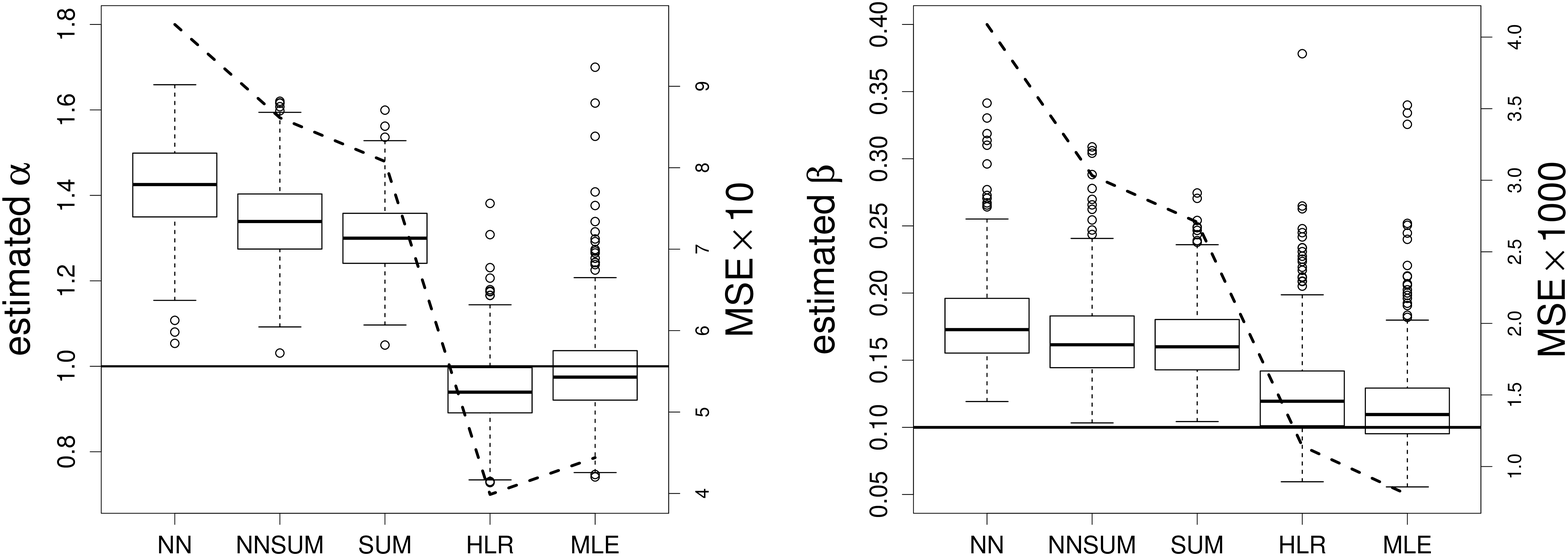}
	\caption{\label{fig:simulation}Two panels showing the boxplot of parameter estimates and mean squared error times 10 (left) or 1000 (right). The solid line is a reference for the true parameter value, and the dash line is the corresponding mean squared error of the $500$ number of estimates in each method. Left: illustration for estimated $\alpha$; Right: illustration for estimated $\beta.$}
\end{figure}

In \S\ref{sec:num}, we calculated the Kullback--Leibler divergence at the true parameter values. In this section, we generate $n=900$ observations with parameters  $\alpha=1, \beta=0.1, \nu=0.5$ and $\tau^2=0.15.$ We run the optimization for $\alpha,\beta,\tau^2$ while fixing $\nu$ at the true value and obtain the estimates of $\alpha,\beta, \tau^2$ by maximizing the approximated likelihoods with $r=2$. We repeat the estimates procedure $500$ times and the boxplots of $\alpha$ and $\beta$ are shown in Fig.~\ref{fig:simulation}.
We see that the estimates obtained by the hierarchical low rank approximation method have the smallest mean squared error among the 3 approximation methods and are close to the exact maximum likelihood estimation.

\section{Application}\label{sec:app}
\subsection{Dataset description}
\begin{figure}[ht]
	\centering
	\includegraphics[width=0.9\textwidth]{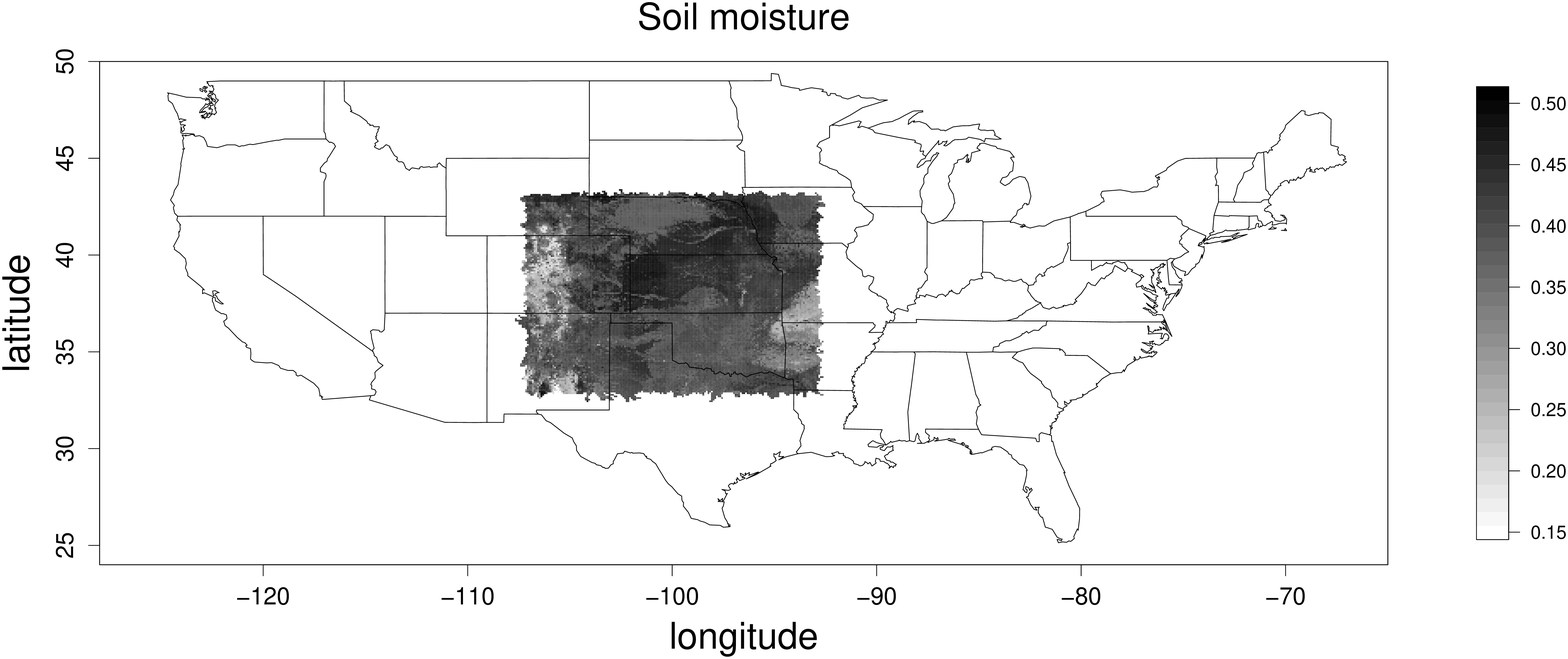}
	\caption{\label{fig:data} Soil moisture (unit: percentage) at the top layer of the Mississippi basin, U.S.A. on January $1^\textrm{st}$, 2014. }
\end{figure}

In this section, we apply our method to modeling soil moisture, a key factor in evaluating the state of the hydrological process, including runoff generation and drought development. We consider high-resolution daily soil moisture data at the top layer of the Mississippi basin, U.S.A., on January $1^\textrm{st}$, 2014 \citep{Chaney:Metcalfe:Wood:review}.
The spatial resolution is of $0.0083$ degrees. The grid consists of $1830\times1329=2,432,070$ locations with $2,153,888$ observations and $278,182$ missing values. The illustration of the data is shown in Fig.~\ref{fig:data}.

We know that a one-degree difference in latitude along any longitude line is equivalent to $111$ km; however, the distance of one-degree difference in longitude depends on the corresponding latitude. As the range of the latitude in this region is relatively small, for simplicity, we use the distance of one-degree difference in longitude at the center location of the region to represent all others, which is $87.5$ km; namely, in this region, $1^\circ$ in latitude is $111$ km and $1^\circ$ in longitude is $87.5$ km.

To understand the structure of the day's soil moisture, we fit a Gaussian process model with a Mat\'ern covariance function.
From all the locations, we randomly pick $n=2,000,000$ points, which are irregularly spaced, to train our model. To assess the quality of our model, the fitted models can be used to predict part of the left out observations.

\subsection{Estimation and prediction}\label{subsec:estimation}
\begin{figure}[ht]
	\centering
	\includegraphics[width=0.8\textwidth]{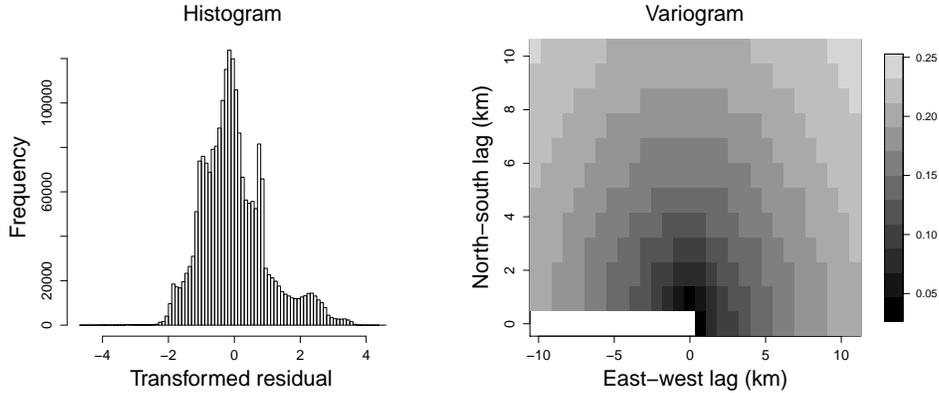}
	\caption{\label{fig:histNvario}Left: the histogram of the transformed residuals; Right: the image plot of the empirical variogram at different distances and along different directions.}
\end{figure}

To use a Gaussian process model, we first fit a linear model to the longitude and latitude as the covariates to the soil moisture. After fitting, we find the negatively skewed residuals, hence we apply a logarithm transformation with some shift.
The histogram of the transformed residual is shown in the left panel of Fig.~\ref{fig:histNvario}, which does not show strong departure from Gaussianity. To examine the isotropy of this process, we calculate the directional empirical variograms as illustrated in the right panel of Fig.~\ref{fig:histNvario}. We see the variograms on the circle with the same radius to the origin have similar values, suggesting that it is reasonable to assume an isotropic model.

Let $z(s)$ denote the transformed residual and the region $D$ be the set of the selected locations, then the proposed Gaussian process model here is
$\{z(s):s\in
D\subset\mathbb{R}^d\}
\sim
\textsc{GP}(0,C(h;\theta)).$ We choose three different covariance functions: the exponential, which has the smoothness parameter $\nu=0.5$; the Whittle, which has $\nu=1$; and the Mat\'ern covariance function, which has an unknown $\nu$.
The formula is given in equation~\eqref{eq:matern}.
Given that the $2,000,000$ observations follow $Z\sim N(0,\Sigma(\theta))$,   $\Sigma(\theta)$ is the 2 million by 2 million variance-covariance matrix, obtained from the chosen covariance function. We use nearest neighbors and hierarchical low rank approximation methods with rank $r=60$ to get the approximated likelihood and then obtain the parameter estimates. The results are shown in Table~\ref{tab:estimate}.
The Mat\'ern covariance model is more flexible by allowing to estimate $\nu$. The estimated $\nu$ in the 
Mat\'ern covariance model by both methods is smaller than $0.5$, and the estimated $\beta$ has the largest value. It suggests a rougher process with a larger dependence range compared to the estimated exponential covariance model.
The last row of Table~\ref{tab:estimate} shows the values of log-likelihood per observation. For each given covariance model, the likelihood with parameters estimated by the hierarchical low rank approximation method is always larger than that by the nearest neighbors method.
Among different covariance models, the likelihood with Mat\'ern covariance is the largest.

\begin{table}[h]
	\def~{\hphantom{0}}
	\centering
	\caption{Parameter estimation results}
\small{	
		\begin{tabular}{l|ccc|ccc}
			\hline
			& \multicolumn{3}{c|}{Nearest neighbors}& \multicolumn{3}{c}{Hierarchical low rank approximation} \\
			& Exponential & Whittle & Mat\'ern  &  Exponential & Whittle & Mat\'ern \\\hline
			Estimated~ $\alpha$ 
			& $~~1.0073~$ & $~~0.9787~$ & $~~1.0597~$ 
			& $~~1.0065~$ & $~~0.9789~$ & $~~1.0539~$\\
			Estimated~ $\beta$ (km) 
			& $~21.6115~$ & $~~5.9316~$ & $222.6545~$ 
			& $~21.2944~$ & $~~5.8216~$ & $178.2051~$\\
			Estimated~ $\tau^2$ 
			& $~~0.0107~$ & $~~0.0013~$ & $~~0.0000~$ 
			& $~~0.0096~$ & $~~0.0012~$ & $~~0.0001~$\\
			Estimated~ $\nu$ 
			& $~~0.5000~$ & $~~1.0000~$ & $~~0.2079~$ 
			& $~~0.5000~$ & $~~1.0000~$ & $~~0.2214~$\\
			\hline
			log-likelihood$/n$
			& $-0.1042~$ & $-0.1417~$ & $-0.0852~$ 
			& $-0.0941~$ & $-0.1308~$ & $-0.0761~$\\
			\hline
		\end{tabular}
		}
		\label{tab:estimate}
	\end{table}

The size of the problem in this application is in the millions, a dataset which is far beyond the ability of classic analysis methods. 
However, nearest neighbors and hierarchical low rank approximation methods can evaluate the approximated likelihood at each iteration in the optimization procedure within 5 and 14 minutes, respectively. The fast computation makes it highly practical for applying the proposed methods to a large real-world spatial dataset problem. The experiment is performed with the Intel Xeon E5-2680 v3@2.50GHz processor.

Next, we use the fitted Mat\'ern model by the hierarchical low rank approximation method to predict soil moisture at the $1000$ left out locations by kriging, which is known to provide the best linear unbiased prediction as well as the prediction standard errors \citep{Cressie:1993}. However, the problem here is of size $n=2,000,000$, hence kriging cannot be employed directly, because it involves a linear solve of size $n$ \citep{Furrer:Genton:Nychka:2006}.
In fact, the proposed methods in this paper can be adopted for approximating kriging equations as well. But for the purpose of validating the fitted model, we explore the exact computation method by treating the irregularly spaced data as observations on a finer regular grid with missing values. The resulting covariance matrix has a block Toeplitz Toeplitz block structure, which can be embedded in a block circulant circulant block matrix \citep{Kozintsev:1999}. Then kriging can be done by fast Fourier transformation. More details can be found in \citet{Chan:Ng:1996}. The mean squared prediction errors over the $1000$ validation locations is $4.53\times10^{-5},$ which is notably small.

\section{Discussion}\label{sec:end}
The implementation in this paper was done with a single-thread program, however as aforementioned in \S\ref{subsec:comp}, computation in each hierarchy can be paralleled, which would reduce the computation time dramatically and make applications even more practical. 
The proposed method can be also extended to more complicated settings. For example, although the rank was fixed to the same in each hierarchy, it can be chosen flexibly in accordance with the number of ``past" observations that are involved in the hierarchy, which, we believe, would give a better approximation. 
Moreover, for prediction problems, the proposed method can be further investigated to approximate kriging equations for large irregularly spaced spatial datasets.

\newpage
\section*{Appendix}
	Recall that the dimension of $V^\textsc{H}_{jj}$ is $mr\times mr$, $V^\textsc{N}_{jj}$ is $r\times r$, $A_{j,r}$ is $j\times r$, $A_{j,mr}$ is $j\times mr$, and $\Sigma_{jj}$ is $j\times j.$ Define $B$ to be the $mr\times r$ matrix by keeping the $mr$ selected rows from $A_{j,r}$, or $B=A_{j,mr}\trans A_{j,r}$. Let $M$ denote $A_{j,mr}\trans\Sigma_{jj}A_{j,mr}$.
The proof of Theorem~\ref{theorem} is as follows.

{\bf Proof of Theorem~\ref{theorem}}
Since the equation,
	\[
	\begin{array}{rl}
		&\|A_{j,mr}V^\textsc{H}_{jj}A_{j,mr}\trans-\Sigma_{jj}\|^2_F
		-
		\|A_{j,r}V^\textsc{N}_{jj}A_{j,r}\trans-\Sigma_{jj}\|^2_F\\
		=&
		\|A_{j,mr}\trans( A_{j,mr}V^\textsc{H}_{jj}A_{j,mr}\trans-\Sigma_{jj})A_{j,mr}\|^2_F
		-
		\|A_{j,mr}\trans(
		A_{j,r}V^\textsc{N}_{jj}A_{j,r}\trans-\Sigma_{jj})A_{j,mr}\|^2_F\\
		=&\|V^\textsc{H}_{jj}-A_{j,mr}\trans\Sigma_{jj}A_{j,mr}\|^2_F
		-
		\|BV^\textsc{N}_{jj}B\trans-A_{j,mr}\trans\Sigma_{jj}A_{j,mr}\|^2_F\\
		=&\|V^\textsc{H}_{jj}-M\|^2_F\
		-\|BV^\textsc{N}_{jj}B\trans-M\|^2_F,\\
	\end{array}
	\]
it suffices to show,
	\[\|V^\textsc{H}_{jj}-M\|_F\
	\leq\|BV^\textsc{N}_{jj}B\trans-M\|_F.\]
Noting that $V^\textsc{H}_{jj}=P_jL_{j}P_j\trans+\epsilon_j^2I_{mr},$ we have $\|V^\textsc{H}_{jj}-M\|_F=\|P_jL_{j}P_j\trans-(M-\epsilon_j^2I_{mr})\|_F$. 
Since $\epsilon_j^2<(\lambda_r+\lambda_{mr})/2$, we know that the eigenvalues of $M-\epsilon_j^2I_{mr}$ satisfy $\lambda_1-\epsilon_j^2\geq\lambda_2-\epsilon_j^2\geq\cdots\geq\lambda_{r}-\epsilon_j^2$ and $|\lambda_{r}-\epsilon_j^2|\geq\max^{mr}_{k=r+1}(|\lambda_k-\epsilon_j^2|)$.
Thus, $|\lambda_1-\epsilon_j^2|\geq|\lambda_2-\epsilon_j^2|\geq\cdots\geq|\lambda_{r}-\epsilon_j^2|\geq\max^{mr}_{k=r+1}(|\lambda_k-\epsilon_j^2|)$. By the construction of $P_j$ and $L_j$, and Eckart-Young-Mirsky theorem \citep{Eckart:Young:1936,Mirsky:1960}, we know,
	\[
	\|P_jL_{j}P_j\trans-(M-\epsilon_j^2I_{mr})\|_F
	=\inf_{\hbox{rank}(X)\leq r}\|X-(M-\epsilon_j^2I_{mr})\|_F.
	\]
Noting that the rank of $BV^\textsc{N}_{jj}B\trans$ is $r$, we have $\|V^\textsc{H}_{jj}-M\|_F=\|P_jL_{j}P_j\trans-(M-\epsilon_j^2I_{mr})\|_F\leq\|BV^\textsc{N}_{jj}B\trans-(M-\epsilon_j^2I_{mr})\|_F=\|(BV^\textsc{N}_{jj}B\trans-M)-\epsilon_j^2I_{mr}\|_F.$
It is easy to observe that the diagonal elements of $BV^\textsc{N}_{jj}B\trans-M$ is non-positive, thus $\|(BV^\textsc{N}_{jj}B\trans-M)-\epsilon_j^2I_{mr}\|_F\leq\|BV^\textsc{N}_{jj}B\trans-M\|_F.$ Then $\|V^\textsc{H}_{jj}-M\|_F\
\leq\|BV^\textsc{N}_{jj}B\trans-M\|_F.$
This completes the proof.	

\bibliographystyle{agsm}
\bibliography{reference}

\end{document}